# An Advanced Atmospheric Dispersion Corrector: The Magellan Visible AO Camera

Derek Kopon[*a], Laird M. Close[a], Victor Gasho[a]
[a]CAAO, Steward Observatory, University of Arizona, Tucson AZ USA 85721

## ABSTRACT

In addition to the BLINC/MIRAC IR science instruments, the Magellan adaptive secondary AO system will have an EEV CCD47 that can be used both for visible AO science and as a wide-field acquisition camera. The effects of atmospheric dispersion on the elongation of the diffraction limited Magellan adaptive optics system point spread function (PSF) are significant in the near IR. This elongation becomes particularly egregious at visible wavelengths, culminating in a PSF that is 2000μm long in one direction and diffraction limited (30-60 μm) in the other over the wavelength band 0.5-1.0μm for a source at 45° zenith angle. The planned Magellan AO system consists of a deformable secondary mirror with 585 actuators. This number of actuators should be sufficient to nyquist sample the atmospheric turbulence and correct images to the diffraction limit at wavelengths as short as 0.7μm, with useful science being possible as low as 0.5μm. In order to achieve diffraction limited performance over this broad band, 2000μm of lateral color must be corrected to better than 10μm. The traditional atmospheric dispersion corrector (ADC) consists of two identical counter-rotating cemented doublet prisms that correct the primary chromatic aberration. We propose two new ADC designs: the first consisting of two identical counter-rotating prism triplets, and the second consisting of two pairs of cemented counter-rotating prism doublets that use both normal dispersion and anomalous dispersion glass in order to correct both primary and secondary chromatic aberration. The two designs perform 58% and 68%, respectively, better than the traditional two-doublet design. We also present our design for a custom removable wide-field lens that will allow our CCD47 to switch back and forth between an 8.6" FOV for AO science and a 28.5" FOV for acquisition.

## 1. INTRODUCTION: MAGELLAN VISIBLE AO

The excellent seeing conditions at the Magellan site frequently provide $r_o$ as high as 20 cm at 0.55 μm. Because of this, we expect that at λ~ 0.9 μm there will be AO correction on bright stars and that moderate Strehls will be possible in the I and z bands on good nights. The resulting angular resolutions of such CCD images will be a spectacular 20-30 mas (although the corrected FOV will typically be limited by the isoplanatic angle to less than 8.5"). To maximize this Vis AO science without impacting our Mid-IR AO mode we have designed an AO science CCD ready at first light to capitalize on these high resolutions (2x better than the best HST ACS images). Our optical Strehl estimates (Esposito et al. 2008) may be optimistic (as these AO Strehl estimates tend to be). However, there is reason to expect that in the z, I, and R bands Strehls will be at least greater than 15%, 10%, and 5%, respectively, when $r_o$>20 cm for stars brighter than V=10. While these Strehls are low compared to what will be achieved at 10 μm, there is still a large body of science that can be done at low Strehl. Most current ~200 actuator 8-10m AO systems do not achieve Strehls much higher than 2% in the I band (0.85 μm). If we estimate no better control than these current systems, and note that our fitting error is a factor of 2x $rad^2$ better, then it is clear that our Strehls with bright stars (fitting error limited) will trend towards I band Strehls of 16%.

To take full advantage of the periods in a night when the seeing is 0.5" requires the Vis AO camera to "always be ready". Our Vis AO camera is conveniently integrated into the WFS stage. We can select a beam splitter to steer ~50% of the WFS visible light into the Vis AO camera with 8.5 mas pixels. We have also designed a removable wide-field lens that can change the FOV of the Vis AO camera by a factor of x3 in order to switch back and forth between a narrow-field science camera and a wide-field acquisition camera.

The AO system currently being built for the Magellan telescope (Figures 1a and 1b) consists of an adaptive secondary mirror (ASM) built by the University of Arizona mirror lab and a pyramid wavefront sensor (PS) built by the Osservatoria Astrofisico di Arcetri (Esposito et al. these proceedings, paper 7015-229). The ASM is identical in optical prescription to the LBT ASM and will use all the same hardware and control software (see these proceedings for more on the LBT AO systems; Riccardi et al. 2008: paper 7015-37). The primary infrared science camera is BLINC/MIRAC4, which will receive IR light from a dichroic beam splitter. Visible light reflected by the dichroic will be sent to an optical bench (hereafter called the W-unit) containing the PS and a visible (0.5-1.0μm) science CCD. The layout of the W-unit is shown in Figure 2. In this paper we will review the optical design and performance of the ADC and Vis AO science camera. The rest of the system is discussed in Close et al. (these proceedings, paper 7015-33).

---

[*] dkopon@as.arizona.edu

Light entering the W-unit passes through a triplet lens that converts it from a diverging F/16 beam into a converging F/49 beam. This light then passes through an ADC before hitting a beam splitter wheel. Light transmitted through the beam splitter wheel then hits a fast steeling mirror, a K-mirror rerotator, and the double pyramid before being re-imaged on the CCD39 for wavefront measurement. A detailed description of the operation of the PS arm of the W-unit can be found in Esposito et al. 2008

The light reflected from the beam splitter will travel to the CCD47, which will be used as both an acquisition and visible science camera. In science mode, the converging F/49 beam results in a square FOV of 8.6". A removable multi-element lens in front of the CCD47 will be able to speed up the beam and increase the FOV by a factor of three, thereby allowing the CCD47 to be used as an acquisition camera and wide-field science camera.

Our objective for the CCD47 in narrow field mode is diffraction limited image quality over the full 8.6" FOV over the band 0.5-1.0μm out to a Zenith angle of 70°. Meeting this tight performance requirement over a broad band at high zenith angles requires a high performance ADC. The criteria used to evaluate relative performance of various designs is the total rms spot size relative to the spot centroid for six different wavelengths spanning the 0.5-1.0μm range in increments of 0.1μm. The Zemax "Atmospheric" surface was used to simulated the atmospheric dispersion with estimated Magellan site parameters (humidity, temperature, etc.).

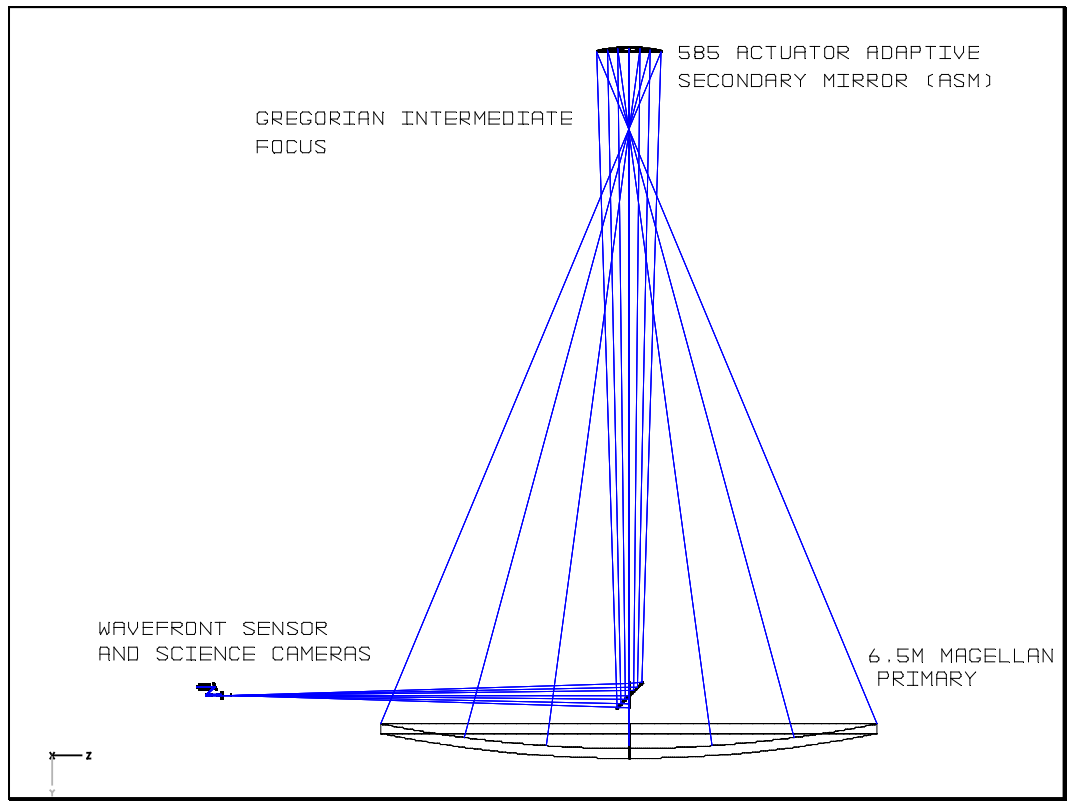

**Figure 1a:** Raytrace of the Magellan 6.5m telescope.

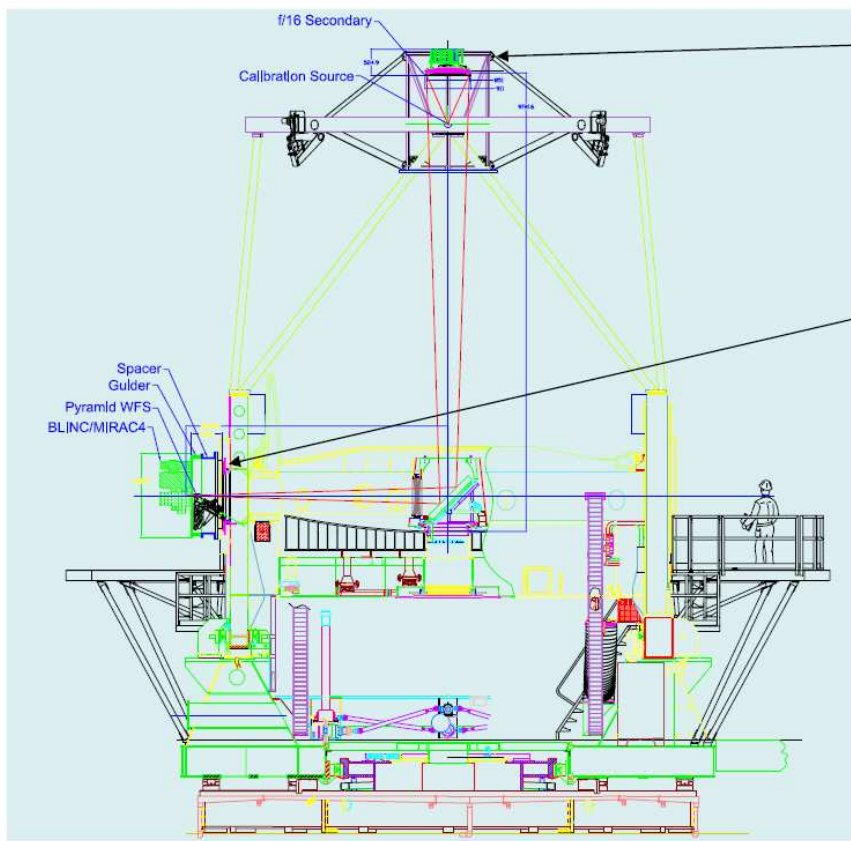

**Figure 1b:** Schematic layout of the Magellan telescope, the adaptive secondary mirror, and the science instruments. Figure taken from Close, et al. (2008; these proceedings).

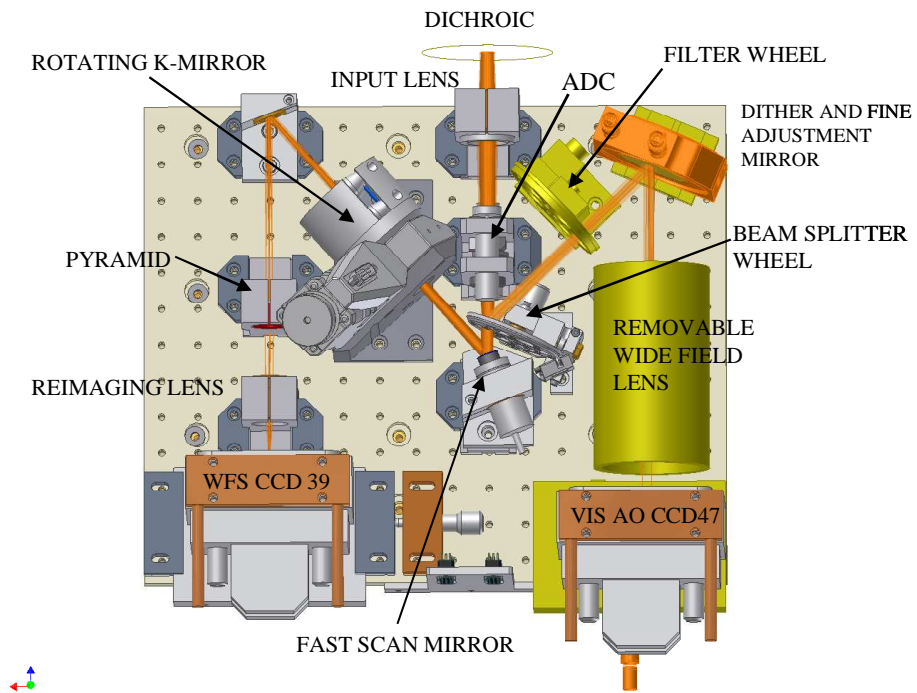

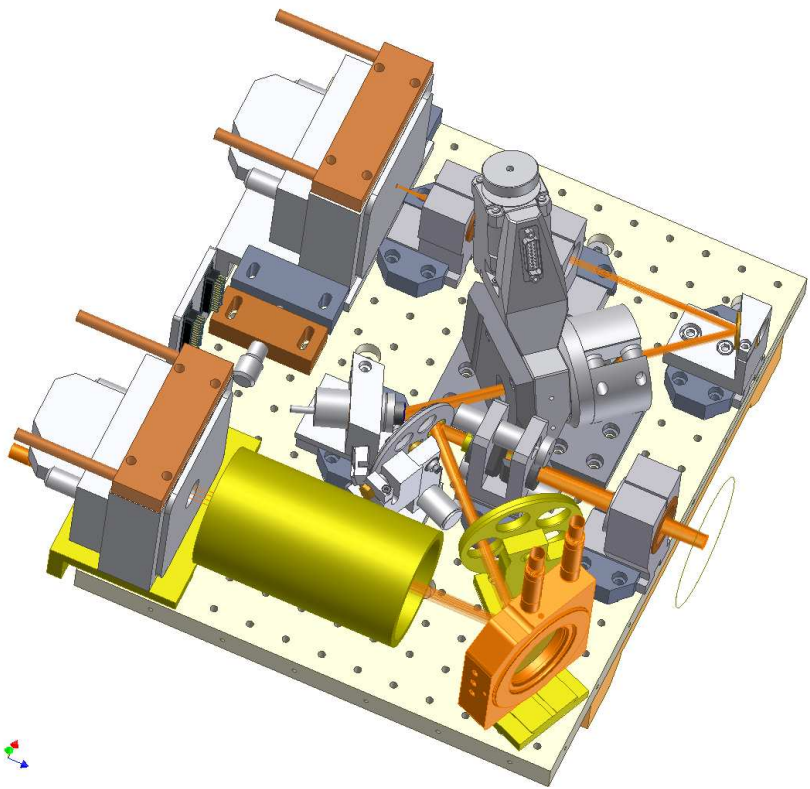

**Figure 2**: The W-unit layout. The incoming beam passes through the ADC before hitting a beam splitter. The transmitted beam goes to the pyramid wavefront sensor and the reflected beam goes to the CCD47 visible science/acquisition camera.

## 2. ATMOSPHERIC DISPERSION CORRECTION

### 2.1 The 2-Doublet Design

Most ADCs designed and built to date consist of two identical counter-rotating prism doublets (often referred to as Amici prisms) made of a crown and flint glass. The indices of the two glasses are matched as closely as possible in order to avoid steering the beam away from its incident direction. The wedge angles and glasses of the prisms are chosen to correct primary chromatic aberration at the most extreme zenith angle. By then rotating the two doublets relative to each other, an arbitrary amount of first-order chromatic aberration can be added to the beam to exactly cancel the dispersion effects of the atmosphere at a given zenith angle. The 2-Doublet design evaluated in this paper is that of the Arcetri group, which will be used on the LBT WFS and was originally designed to be diffraction limited over the band 0.6-0.9 μm out to ~65° zenith angle.

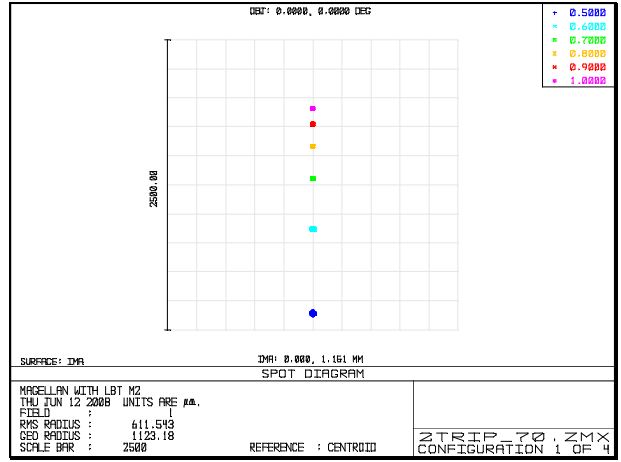

**Figure 3:** Example of the uncorrected atmospheric dispersion at 50° zenith (1.55 airmasses). The diffraction-limited PSF over the 0.5-1.0μm band is 30μm/60μm x 2000μm without ADC correction.

As shown in the spot diagram in Fig. 4, the 2-Doublet design corrects the atmospheric dispersion so that the longest and shortest wavelengths overlap each other, thereby correcting the primary chromatism. Secondary chromatism is not corrected and is the dominant source of error at higher zenith angles. To correct higher orders of chromatism, more glasses and thereby more degrees of freedom are needed.

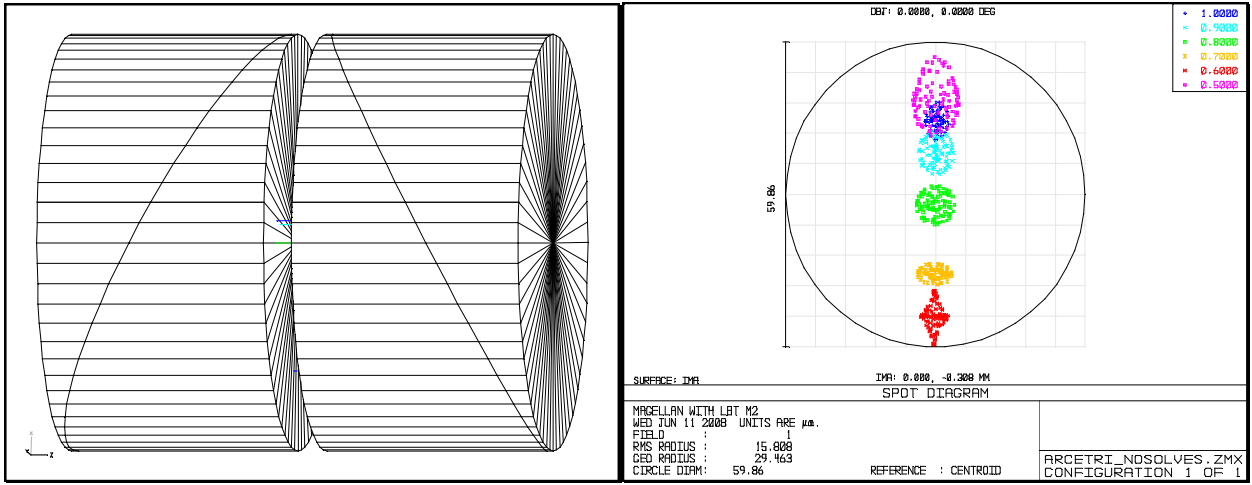

**Figure 4:** The classic 2-doublet ADC design and an AO spot diagram at 50° zenith. Notice that the longest and shortest wavelengths overlap, indicating that primary chromatism is almost perfectly corrected. The circle has a radius of 29.9 μm, which is the diffraction limit at 0.5 μm. Lateral secondary chromatism is the dominant aberration.

## 2.2 The 2-Triplet Design

In our 2-triplet design, a third glass with anomalous dispersion characteristics (Schott's N-KZFS4) is added to the crown/flint pair. Like the doublet, the index of the anomalous dispersion glass was matched as closely as possible to that of the crown and flint. The Zemax atmospheric surface was set to 70 deg zenith and the relative angles of the ADC were set to 180 deg. The wedge angles of the three prisms in the triplet were then optimized to correct both primary and secondary chromatism.

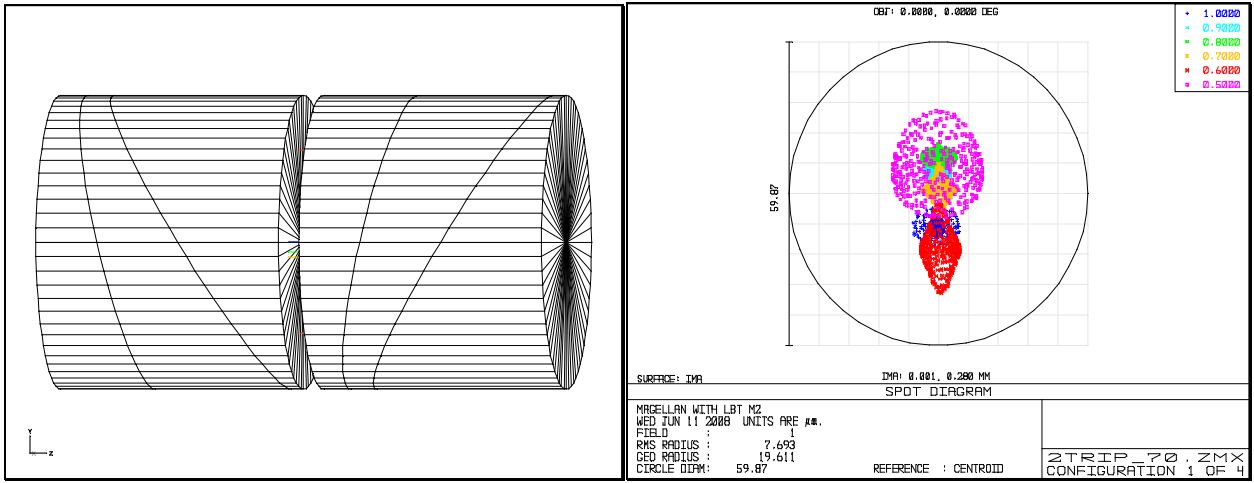

**Figure 5:** Our 2-triplet design at 50 deg zenith (1.55 airmasses). Both primary and secondary chromatism are well corrected, leaving only small amounts of higher order lateral and axial chromatic aberrations.

## 2.3 The 4-Doublet Design

This design philosophy can be carried one step further by adding a forth glass and another rotational degree of freedom in a 4-doublet design. The first and second doublets are identical to each other and the third and forth are identical. A multi-configuration Zemax model was optimized for several hours over the full Zenith range, glass catalogue, and possible wedge angles to generate a design that has better overall performance than either the 2-doublet or 3-triplet design. It was also initially thought that the 4-doublet design would be more impervious to errors in fabrication or in the Zemax model atmosphere than the 2-triplet design, since the two doublet pairs can essentially correct primary and secondary chromatism independently of each other. In contrast, the relative amounts of primary and secondary chromatic correction are fixed by the fabricated wedge angles in the 2-triplet design. However, a tolerance analysis shows that the 2-triplet design can handle small fabrication errors and chromatic fluctuations in the atmosphere due to changes in humidity, temperature, or pressure. Since the 2-triplet design is less complex than the 4-doublet design and has sufficient performance for our needs, it is the baseline design for the Magellan Vis AO system.

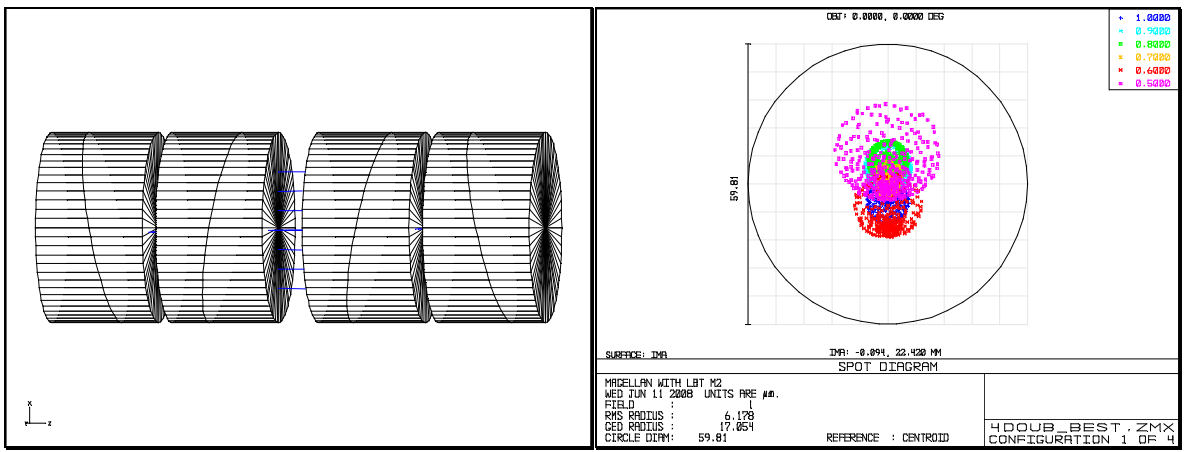

**Figure 6:** The 4-doublet design correcting the 2000um of chromatic aberration at 50 deg zenith to better than 6um rms.

Figure 7 shows the rms spot size over the 0.5-1.0 µm band as a function of zenith angle for the three designs. At the CCD47 in narrow-field science mode, the 0.5 µm diffraction limit is 29.84 µm (radius of first Airy minima). It should also be noted that over a narrower band, such as 0.6-0.9 µm, rms spot size is much smaller: roughly a factor of x2 at most zenith angles. The 0.5 µm spot is usually the biggest contributor to a large spot size at high zenith angle for the three designs.

As Figure 7 shows, the 2-doublet and 2-triplet designs both have sharp spikes in rms spot size at 0° zenith. This is because there is no atmospheric dispersion, so there is no chromatism to cancel the intrinsic chromatic residual of the ADC. The 2-doublet and 2-triplet designs suffer from this effect because they only have two elements. The 4-doublet design does not suffer from this effect because two of the elements can cancel the residual chromatism from the other two. Regardless, any telescope large enough to require a high-performance ADC for visible AO will be alt-az mounted and will not be observing exactly at zenith.

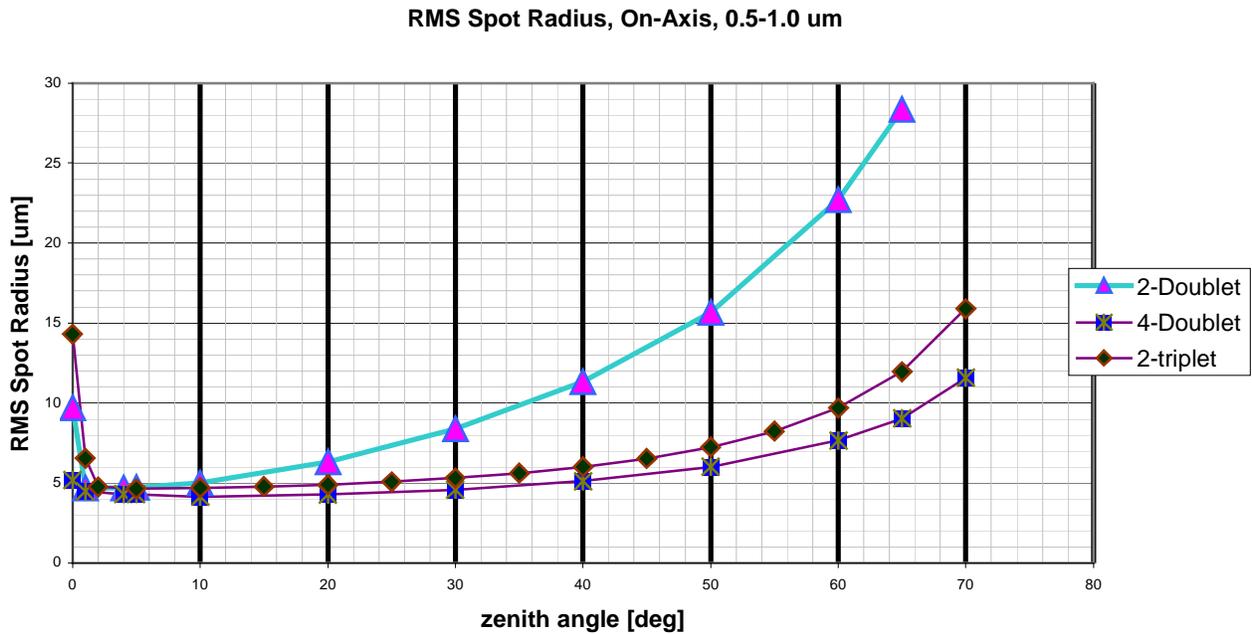

**Figure 7:** RMS spot size vs. zenith angle for the three ADC designs. The 2-triplet ADC is the baseline design for the Magellan AO system due to its relative simplicity and diffraction limited performance down to three airmasses.

### 2.4 Pupil Shear

Because the pyramid sensor operates by imaging the pupil, it is essential that any ADC design not introduce any significant chromatic pupil shearing. Chromatic pupil shear of more that ~10-20% of the 24 µm pixel pitch of the CCD39 could interfere with our wavefront sensing. Chromatic imaging effects cause pupil shear to be weakly dependant on field angle, pupil position, and ADC design, and more strongly dependant on zenith angle. In order to quantify the pupil shear, we traced individual rays over the 0.6-0.9 µm wavefront sensing band both on-axis and at the edge of the field at 20 different pupil positions for each field point for a given zenith angle and ADC design. The shear was then calculated as the maximum distance on the CCD39 between rays of different wavelengths for a given field point and pupil position. The shears calculated with these rays over the two field points and 20 different pupil positions were then averaged and are plotted in Figure 8. The standard deviation of all the shears for a given ADC design and zenith angle was small, typically ~2%. As Figure 8 shows, the shear can become significant at higher zenith angles (50% at 70° zenith). To compensate for this effect, we may need to either use a narrower wavelength band for wavefront sensing, or bin the CCD39 in a coarser sampling mode.

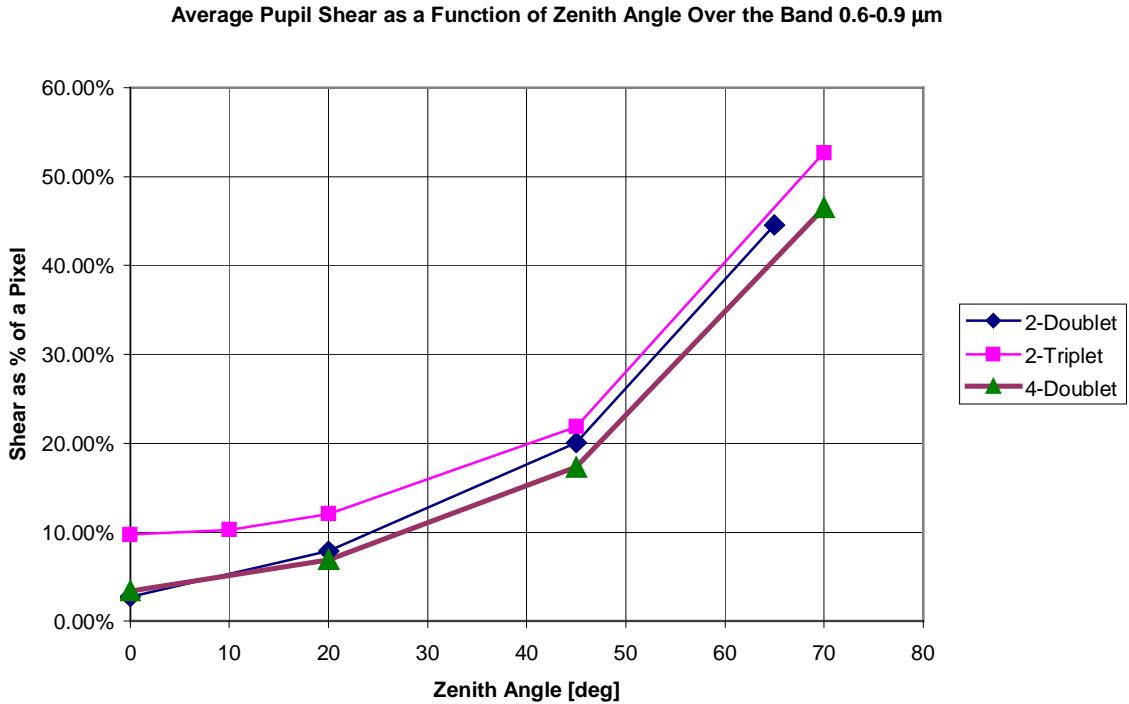

**Figure 8:** Chromatic pupil shear as a function of zenith angle over the 0.6-0.9 µm wavefront sensing band. Each point on the plot is an average of shear values at two different field points and 20 different pupil positions.

## 3. WIDE-FIELD LENS

The EEV CCD47 of the W-unit will serve as both a narrow field (8.6" FOV) science camera for visible AO and a wide field (28.5" FOV) acquisition camera. Because of the fixed position of the CCD47 and the tight geometry of the W-unit board (Figure 2), we were forced to design a removable lens that would fit between the final fold mirror and the focal plane. Typically, a lens of this type is more naturally located near a pupil. However, all of the real estate near the pupil is occupied by the ADC, the beam splitter, and the filter wheel. Because of the large distance from the pupil, our lens elements must have large (3") diameters and higher curvatures than most stock lenses.

Our wide-field design is diffraction limited over the band 0.8-1.0 µm in the inner ~20" diameter circle of the CCD, which is the largest possible isoplanatic patch we would ever expect to have at 1.0 µm. The field over the rest of the CCD outside of this well-corrected inner patch is not diffraction limited, yet is still much better than a seeing-limited spot of 1". By relaxing the image requirements on this outer portion of the field, we were able to generate a relatively inexpensive design that makes use of two stock Melles Griot achromats along with a custom negative doublet. The distortion of the lens is also small, with a maximum of 1.3% at the edge of the field. The wide field lens will be on a movable mount that will lift the lens into and out of the beam without requiring a focus shift at the detector. However, the CCD will be on a small precision translation stage with +/- 5 mm travel in the z-direction in order to allow us to adjust focus. Figure 9 shows the Zemax design of the Vis AO camera in wide-field mode, with the lens elements in place. Figures 10 and 11 show the spot diagrams of the Vis AO camera in both the wide-field mode and narrow-field mode at low zenith angle.

Because any bright sources will immediately saturate the Vis AO camera, we will have a 0.1" chrome dot and a 1.0" chrome dot on a window immediately in front of the CCD47 that will act as neutral density filters. In order to steer the star onto one of these dots, or to look at sources with the CCD47 that are off-axis from the guide star being imaged on the pyramid sensor, the fold mirror immediately prior to the wide-field lens location will be an automated fine-adjustment tip/tilt stage.

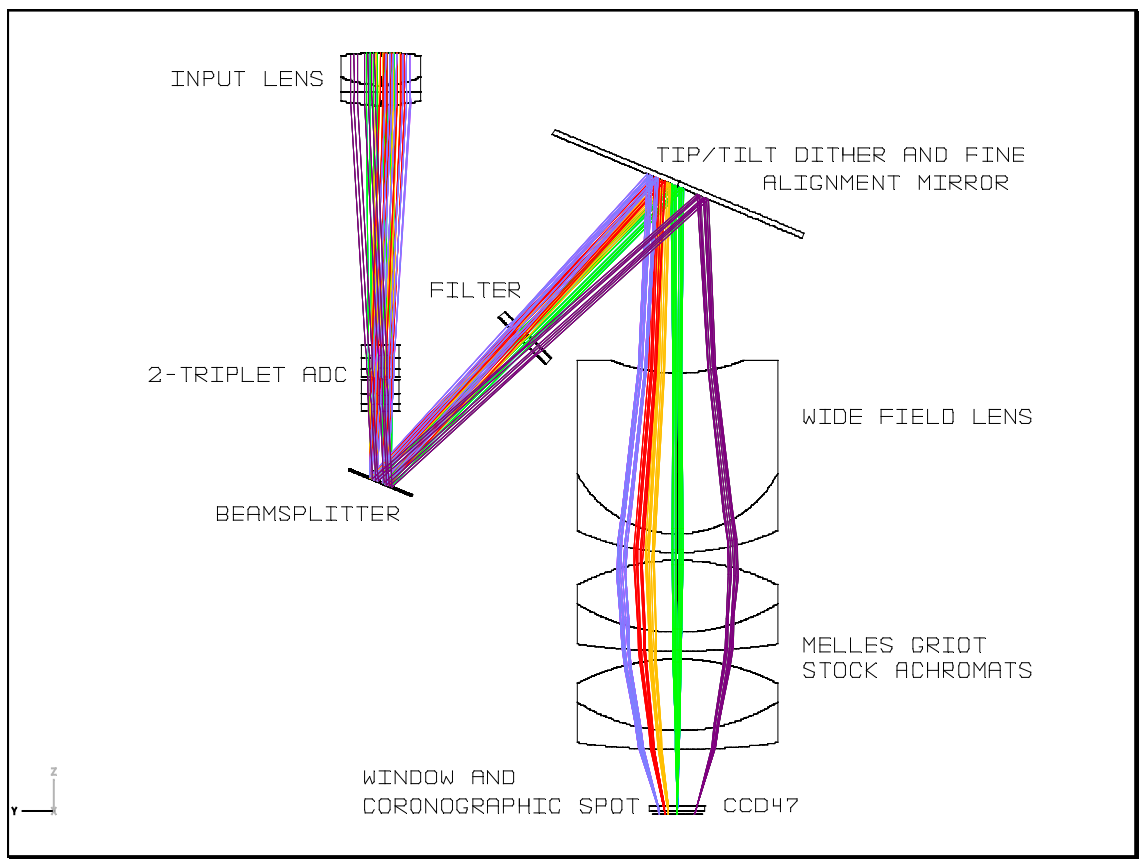

**Figure 9:** The Vis AO camera will have a three element removable wide field lens that can increase the FOV from 8.6" to 28.5". Two of the three elements are identical Melles Griot stock achromats.

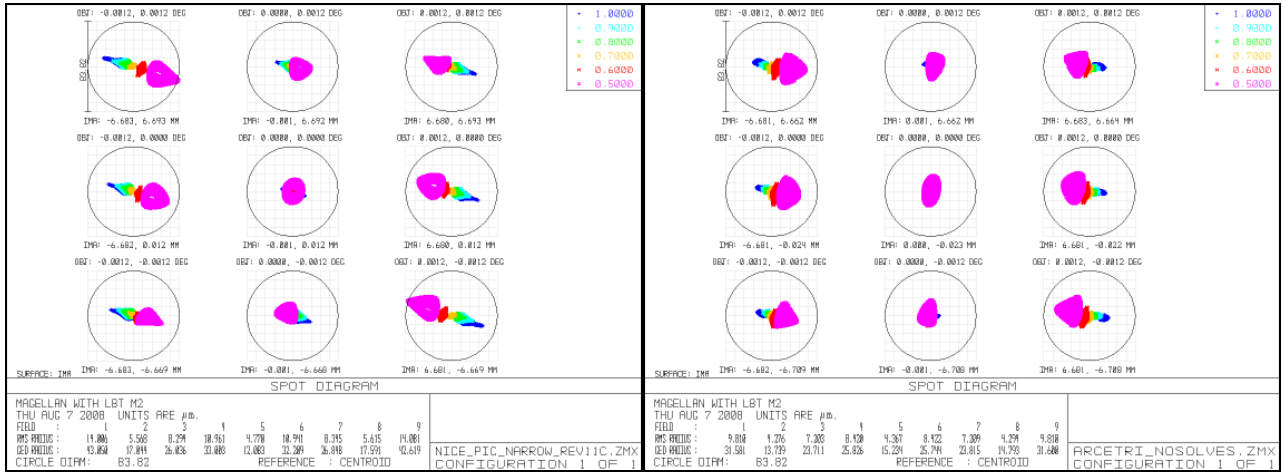

**Figure 10:** The spot diagrams for the Vis AO CCD47 in narrow-field (8.6" FOV) mode. The box to the left contains the center and edge spots produced with the 2-triplet ADC design. The box to the right contains the center and edge spots for the 2-doublet design. The variation in chromatism across the field is a residual caused by the ADCs themselves and does not vary with zenith angle. Both spot diagrams were made at 5° zenith to avoid zenith spike artifacts at 0° zenith. The circle shown is the diffraction limit for 0.7 µm, which is the shortest wavelength at which we expect our AO system to be diffraction limited. While the spot diagrams for the 2-doublet ADC are slightly better than those of the 2-triplet design at this low zenith angle (see figure 7), the 2-triplet design outperforms the 2-doublet design at higher zenith angles and is therefore the design of choice for the Magellan AO system.

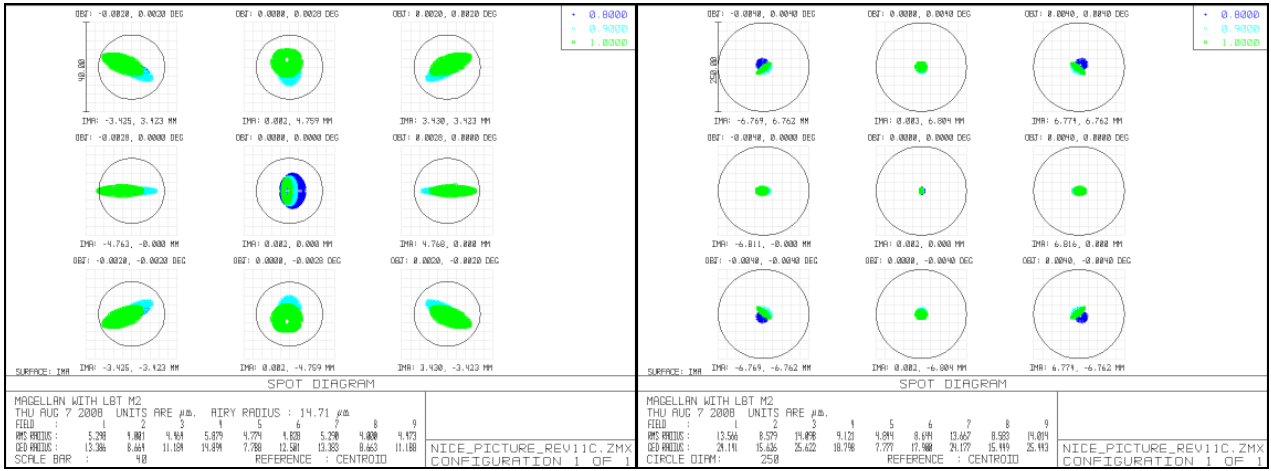

**Figure 11:** The spot diagrams for the CCD47 in wide-field (28.5") mode. The spot diagrams to the left are the center and edges of the inner 20" circular patch, which is the largest possible isoplanatic patch on a good night. The circles are the diffraction limit at 0.8µm. The spots to the right are the center and edges of the entire 28.5" FOV of the CCD47 with a 0.5" diameter (250µm) seeing-limited circle for comparison. These plots were made with the 2-triplet ADC in place at 0° zenith angle.

## 4. CONCLUSION

In this paper we have discussed the design and performance of two novel ADC designs, the 2-triplet and 4-doublet, and compared them with a classical 2-doublet design. The 2-triplet design gives significantly better performance than the 2-doublet design (56% improvement) with only slightly more complexity. The 4-doublet design is only marginally better from a performance perspective, but significantly more difficult to implement, with twice the number of elements that need to be mounted in the tight confines of the W-unit and controlled. We also present our wide-field lens design that allowed our visible science CCD to operate in both narrow (8.6") and wide-field (28.5") modes.

## ACKNOWLEDGEMENTS

This project owes a debt of gratitude to our partners and collaborators. The ASM and WFS could not have been possible without the design work of Microgate and ADS in Italy as well as Arcetri Observatory and the LBT observatory. We would like to thank the NSF MRI and TSIP programs for generous support of this project. We would also like to thank Simone Esposito, Andrea Tozzi, and Phil Hinz for providing the Zemax design of the LBT pyramid WFS, including the Arcetri 2-doublet ADC.